\documentstyle[b98proc,epsfig]{article}

\def\Journal#1#2#3#4{{#1} {\bf #2}, #3 (#4)}

\newcommand{\slsha}[1]{\; {\not {\!\!\!\: #1}}}
\begin{document}

\title{
PROTONINDUCED PION-, ETA- AND KAON-PRODUCTION AT THRESHOLD
\footnote{Contribution to BARYONS98, Bonn, Germany, September 1998}}

\author{F. KLEEFELD, M. DILLIG}
\address{Institute for Theoretical Physics III, University of Erlangen-N\"urnberg, Erlangen, Germany}


\maketitle

\abstracts{
We present some details of
our recent theoretical investigations on the total
cross section of the reactions $pp\rightarrow pp\,\pi^0$, 
$pp\rightarrow pp\,\eta$ and $pp\rightarrow p\Lambda K^+$
at threshold.
Applying standard nonrelativistic perturbation theory within the framework of
meson exchange models (MEC) to all of the investigated processes we observe that theory
is still far--off from a quantitative understanding of the existing data, although
the qualitative behaviour of the threshold cross sections can be reproduced.
The often quoted $\eta^2$--dependence of the $pp\rightarrow pp\,\pi^0$
cross section at threshold is due to a calculational mistake of Koltun and
Reitan (1966) and not based on theoretical grounds, where threebody phasespace
behaviour would yield $\eta^4$, which can be affected by final state and short
range effects.
To respect the high momentum transfers for 
$pp\rightarrow pp\,\eta$ and $pp\rightarrow p\Lambda K^+$ relativistic
models were developed. Hereby the $\eta$--production still is based on 
MEC, while the $K^+$ production was investigated in 
the quark-gluon-picture based on the Bethe--Salpeter formalism.
}
\section{Final state interactions in the $pp\,\pi^0$ and $pp\,\eta$--system}
The energy dependence of the cross section of $pp\rightarrow pp\,\phi$  
($\phi\in{\pi^0,\eta}$) at threshold is determined by final state modified
phasespace integrals$\;$\cite{kle1}.
A simple example is the shape independent effective range expansion 
in the $pp$--final state
($s_1 := (p^\prime_1 + p^\prime_2)^2$,
$\kappa:= \sqrt{\lambda(s_1,m^2_p,m^2_p)}/(4m_p)$, $s=P^2$):
\begin{eqnarray}
\lefteqn{
 R^{\; \scriptsize \mbox{FSI}}_3 (s \; ; \; m^2_p, m^2_p, m^2_{\phi} ) 
 \approx
\int \! \frac{d^{\,3} p_{1^\prime}}{
2 \, \omega_{p} (|\vec{p}_{1^\prime}|)} \;
\frac{d^{\,3} p_{2^\prime}}{
2 \, \omega_{p} (|\vec{p}_{2^\prime}|)} \;
\frac{d^{\,3} k_\phi}{
2 \, \omega_{\,\phi} (|\vec{k}_\phi|)} } \cdot \nonumber \\
 & \cdot & \delta^4 (p_{1^\prime} + p_{2^\prime} + k_\phi - P\, ) 
 \left| 
  - \, \frac{\mbox{N}_\phi}{a_{pp}}  
 \frac{1}{(- \, 1/a_{pp} \; + \;  r_{pp} 
\, \kappa^2/2
 - i \kappa )}  
 \, \right|^2 \; \alpha \;\; \eta^4_\phi \qquad
\end{eqnarray}  
The dimensionless normalization $\mbox{N}_\phi$ is an open question in nearly
all theoretical models in the literature which prevents them from
being quantitative.  
\section{The reaction $pp\rightarrow pp\,\pi^0$ at threshold}
Applying a nonrelativistic meson exchange model$\;$\cite{kle1,kle2} to $pp\rightarrow pp\pi^0$ at
threshold we are able to describe the total cross section qualitatively upto $\eta_\pi\approx
0.3$. The model contains contributions from IA, offshell 
rescattering, heavy meson exchanges, resonant S-- and P--waves, 
recoil currents and $\pi\pi$--box--diagrams.   
\section{The reaction $pp\rightarrow pp\,\eta$ at threshold} 
The total cross section of $pp\rightarrow pp\,\eta$ at threshold has been
calculated within 
a nonrelativistic and a relativistic meson--exchange model$\;$\cite{kle1,kle2} 
(exciting the $S^-_{11}(1535)$) which is based on the following interaction Lagrangian:
\begin{eqnarray} \lefteqn{{\cal L}_{int} (x) = 
- \, g_{\delta NN}
\bar{N} \vec{\tau} \cdot {\vec{\phi}}_{\delta} \, N  
- \, g_{\sigma NN}
\bar{N}  \phi_{\sigma} \, N  
- i \, g_{\pi NN}
\bar{N} \gamma_5 \, \vec{\tau} \cdot {\vec{\phi}}_{\pi} \, N \, -} \nonumber \\ 
& - & i \, g_{\eta NN}
\bar{N} \gamma_5 \, \phi_{\eta} \, N  
- \, g_{\rho NN} \,
\bar{N} \, \vec{\tau} \cdot [ \gamma_\mu \, {\vec{\phi}}^{\,\mu}_{\rho}   
 + \frac{K_\rho}{2 m_N} \,  
\sigma_{\mu\nu} \, \partial^\mu {\vec{\phi}}^{\,\nu}_{\rho} ] 
\, N \, - \nonumber \\
 & - &
 g_{\omega NN} \,
\bar{N} \, [ \gamma_\mu \, \phi^{\,\mu}_{\omega}   
 + \frac{K_\omega}{2 m_N} \,  
\sigma_{\mu\nu} \, \partial^\mu \phi^{\,\nu}_{\omega} ] 
\, N  
+ \Big[ \, 
i \, g_{\delta NS_{11}} \;
\bar{N}_{S_{11}} \gamma_5 \, \vec{\tau} \cdot {\vec{\phi}}_{\delta} \; N \, + 
\nonumber \\
 & + &
i \, g_{\sigma NS_{11}} \;
\bar{N}_{S_{11}} \gamma_5 \, \phi_{\sigma} \; N   
- g_{\pi NS_{11}} \;
\bar{N}_{S_{11}} \vec{\tau} \cdot {\vec{\phi}}_{\pi} \; N  
- g_{\eta NS_{11}} \;
\bar{N}_{S_{11}} \phi_{\eta} \; N \, +  
 \nonumber \\
 & + & 
i \, g_{\rho NS_{11}} \,\bar{N}_{S_{11}} \gamma_\mu \gamma_5 
\, \vec{\tau} \cdot {\vec{\phi}}^{\,\mu}_{\rho} N   
+ i \, g_{\omega NS_{11}} \,\bar{N}_{S_{11}} \gamma_\mu \gamma_5 
\, \phi^{\,\mu}_{\omega} N   
 + h.c. \Big] 
\end{eqnarray}
Following Watson and Migdal we multiplicatively separate the long-ranged 
final state interactions from the T--matrix. The short ranged (approximately
constant) production amplitude
is set to its threshold value. The resulting cross section of the relativistic
model is ($m_N \simeq (m_p+m_n)/2$):
\begin{eqnarray}
\lefteqn{
\sigma_{pp\rightarrow pp\,\eta} (s) \quad \simeq
\quad 
\frac{1}{2\,!} \; \frac{1}{(2\pi )^5} \;\;  
\frac{R^{\, \mbox{\scriptsize FSI}}_3 (s\; ; \; m^2_p, m^2_p,m^2_\eta )}{
2\; \sqrt{\lambda (s\; ; \; m^2_p, m^2_p )}}
 \;\;
m_\eta \; (m_\eta + 4\, m_{\scriptscriptstyle N} \,) \; \cdot} \nonumber \\
 & \cdot & \Big| \;
  2\, m_{\scriptscriptstyle N} \, 
 \; [ \,
(\, X_{\,\delta } + X_{\,\sigma } \, ) \, (m_{\scriptscriptstyle N} + m_\eta ) - 
(\, X_{\,\pi } + X_{\,\eta } \, ) \, (m_{\scriptscriptstyle N} - m_\eta ) \, +
\nonumber \\ 
 & & + \,
(\, Y_{\,\delta } + Y_{\,\sigma } - Y_{\,\pi } - Y_{\,\eta } \, ) \, (m_{\scriptscriptstyle N} + m_\eta ) + 
 M_\delta + M_\sigma - M_\pi - M_\eta
 \, ]
 \, - \nonumber \\
 & & - \, X_\rho \; m_{\scriptscriptstyle N} \;
 [ \,
 4 \, (m_\eta - 2\, m_{\scriptscriptstyle N}) +
 K_\rho \, (5 m_\eta - 4\, m_{\scriptscriptstyle N})
 \, ] \; + \nonumber \\
 & & + \,
 [ \,
Y_{\,\rho } \, (m_{\scriptscriptstyle N} + m_\eta ) + 
 \tilde{M}_\rho 
 \, ]
 \; [\, K_{\rho} 
 \; (m_\eta - 4\, m_{\scriptscriptstyle N})
 - 8\, m_{\scriptscriptstyle N} \, ]  - \nonumber \\
 & & - \, X_\omega \; m_{\scriptscriptstyle N} \;
 \left[ \,
 4 \, (m_\eta - 2\, m_{\scriptscriptstyle N}) +
 K_\omega \, (5 m_\eta - 4\, m_{\scriptscriptstyle N})
 \, \right] \; + \nonumber \\
 & & + \,
 [ \,
Y_{\,\omega } \, (m_{\scriptscriptstyle N} + m_\eta ) + 
 \tilde{M}_\omega 
 \, ]
 \; [\, K_{\omega} 
 \; (m_\eta - 4\, m_{\scriptscriptstyle N})
 - 8\, m_{\scriptscriptstyle N} \, ] \; 
 \Big|^{\, 2} 
\end{eqnarray}
with$\;$\cite{kle4} ($\phi\in \{\delta,\sigma,\pi,\eta,\rho,\omega\}$)
($M_{S_{11}}:=m_{S_{11}}- i \,\Gamma_{S_{11}}/2$)($D_\phi (q^2):=(q^2-m_\phi^2)^{-1}$)
($q^2:=- m_p m_\eta$, $p^2:=m_p \; (m_p-2 m_\eta)$,
$P^2:=(m_p+m_\eta)^2$):
\begin{eqnarray}
 & & X_{\,\phi }
 :=  
 D_\phi (q^2) \; g_{\, \phi NN} (q^2) \;
 D_{S_{11}} (p^2) \; g^{\, \ast}_{\phi NS_{11}} (q^2) \;  
 g_{\eta NS_{11}} (m^2_{\, \eta}) \nonumber \\
 & & Y_{\,\phi }
 :=  
 D_\phi (q^2) \; g_{\, \phi NN} (q^2) \;
 D^R_{S_{11}} (P^2) 
 \;  
g^{\, \ast}_{\eta NS_{11}^R} (m^2_\eta) \;  
 g_{\Phi NS_{11}^L} (q^2 ) \nonumber \\
 & & M_\phi := X_\phi \, m_{S_{11}} + Y_\phi \, M_{S_{11}} \quad , \quad
 \tilde{M}_\phi := - X_\phi \, m_{S_{11}} + Y_\phi \, M_{S_{11}} \nonumber \\
 & & 
D^R_{S_{11}} (P^2):=(P^2-M^2_{S_{11}})^{-1} \; , \quad
D_{S_{11}} (p^2):=(p^2-m^2_{S_{11}})^{-1} 
\end{eqnarray}
\section{The reaction $pp\rightarrow p\Lambda\,K^+$ at threshold}
To open the door for future investigations we sketch out some details of our
convariant quark--gluon Bethe--Salpeter (BS) model to describe $pp\rightarrow p\Lambda K^+$ at
threshold$\,$\cite{kle3}.
The total cross section of $pp\rightarrow p\Lambda K^+$ is given by:
\begin{eqnarray} \lefteqn{\sigma_{pp\rightarrow p\Lambda K} (s) \quad =
} \nonumber \\
& = & 
\frac{1 }{2 \, \sqrt{\lambda (s; m^2_p, m^2_p)}} 
\int\!\frac{d^{\,3}p_{1^\prime}}{(2\pi )^3\, 2\,\omega_p
(|\vec{p}_{1^\prime}|)}
\frac{d^{\,3}p_{2^\prime}}{(2\pi )^3\, 2\,\omega_\Lambda
(|\vec{p}_{2^\prime}|)}
\frac{d^{\,3}k}{(2\pi )^3\, 2\,\omega_K (|\vec{k} |)}
\nonumber \\
 & & \nonumber \\
 & & 
 (2\pi )^4 \; \delta^4 (p_{1^\prime} + p_{2^\prime} + k - p_1 - p_2) \;
  \frac{1}{4} \; \sum\limits_{s_1,s_2} \; 
 \sum\limits_{s_{1^\prime},s_{2^\prime}} 
 \; {\left| \, T_{fi} \, \right| }^2 \label{xlb1}
\end{eqnarray}
\begin{eqnarray} i \, T_{fi}  
 & = &  <p\,(\vec{p}^{\;\prime}_1,s^\prime_1) 
 \, \Lambda^0 (\vec{p}^{\;\prime}_2,s^\prime_2) \, K^+ (\vec{k})| \,T \,\Big[ 
 \frac{i}{1!} \; {\hat{{\cal L}}}_{int} (0) +
 \frac{i^2}{2!} \; {\hat{{\cal L}}}_{int} (0) \, {\hat{{\cal S}}}_{int}
 + 
\nonumber \\ 
 & + & 
 \frac{i^3}{3!} \; {\hat{{\cal L}}}_{int} (0) 
 \, {\hat{{\cal S}}}_{int}
 \, {\hat{{\cal S}}}_{int}
 + 
\ldots \Big] \, |p\,(\vec{p}_1,s_1)\,p(\vec{p}_2,s_2)>_{c} \label{xlb2} 
\end{eqnarray}
\begin{equation}
{\hat{{\cal L}}}_{int} (x) \simeq 
\sqrt{4\pi\alpha_s} \;\; \bar{\psi} (x) \;
\frac{\lambda_a}{2} \, \slsha{A}_a (x) \; \psi (x) \; , \;
\alpha_s (Q^2) \simeq 4\pi (\beta_0)^{-1} /  
\ln \left(1+\frac{Q^2}{\Lambda_{QCD}^2}\right) 
\end{equation}
(${\hat{{\cal S}}}_{int}:= \int d^{\,4}\!x\; {\hat{{\cal L}}}_{int} (x)$). The in-- and outgoing protons are considered to be bare three--quark objects
dressed by scalar $q\bar{q}$--pairs. In our model we use properly normalised momentum 
space quark and antiquark creation ($b^+,d^+$) and
annihilation ($b,d$) operators 
($\{ b(\vec{p},\alpha), b^+(\vec{p}^{\;\prime},\alpha^\prime)\}=
\delta_{\alpha\alpha^\prime} \; (2\pi)^3 $ $2\,\omega(|\vec{p}\,|) \; \delta^3
(\vec{p}-\vec{p}^{\;\prime})$, \ldots) and Dirac
spinors ($\bar{u}(\vec{p},s) \, u(\vec{p},s^\prime)=2m \;\delta_{ss^\prime}$, 
$\bar{v}(\vec{p},s)$ $v(\vec{p},s^\prime)=-2m \;\delta_{ss^\prime}$)
to express the asymptotic in-- and outgoing
proton state vectors and the outgoing Kaon state vector 
(upto renormalisation constants) in terms of 
corresponding BS--amplitudes 
($s_i,t_i,c_i$ denote spin, flavour and colour):
\begin{eqnarray} \lefteqn{|K^+ (P) > \quad \simeq \quad - \;
\sum\limits_{s_1,s_2} \;
\sum\limits_{t_1,t_2} \;
\sum\limits_{c_1,c_2} 
\int \frac{d^{\, 3}p_1}{(2\pi )^3\, 2 m_1} 
\; \frac{d^{\, 3}p_2}{(2\pi )^3\, 2 m_2} \; 
\int d^{\, 3}x_1 \, d^{\, 3}x_2 
} \nonumber \\
 & \cdot &
 \exp ( \displaystyle -i\, (\vec{p}_1 \cdot \vec{x}_1 + 
 \vec{p}_2 \cdot \vec{x}_2 )) \; \cdot \nonumber \\
 & \cdot & 
 \bar{u}^{\,(1)} (\vec{p}_1,s_1,t_1,c_1) \;\;
 < 0 | \, T \left( \, \psi^{\,(1)} (x_1) \, \bar{\psi}^{\,(2)} (x_2) \, \right) \, |P ,K^+>
\cdot \nonumber \\
 & \cdot & 
 v^{\,(2)} (\vec{p}_2,s_2,t_2,c_2) \; 
 d^+ (\vec{p}_2,s_2,t_2,c_2) \; 
 b^+ (\vec{p}_1,s_1,t_1,c_1) \;
 |0> 
 {\Big|}_{x^0_1\;=\;x^0_2\;=\;0} \qquad
\end{eqnarray}
\begin{eqnarray} \lefteqn{|p^+ (P) > \quad = 
\sum\limits_{s_1,s_2,s_3} \;
\sum\limits_{t_1,t_2,t_3} \;
\sum\limits_{c_1,c_2,c_3} 
\int \frac{d^{\, 3}p_1}{(2\pi )^3 \; 2 \,m_1} 
\; \frac{d^{\, 3}p_2}{(2\pi )^3 \; 2 \,m_2} 
\; \frac{d^{\, 3}p_3}{(2\pi )^3 \; 2 \,m_3}  
} \nonumber \\
 & & \nonumber \\
 & &
\int d^{\, 3}x_1 \; d^{\, 3}x_2 \; d^{\, 3}x_3 \;
 \exp[ -i\, (\vec{p}_1 \cdot \vec{x}_1 + \vec{p}_2 \cdot \vec{x}_2 +
\vec{p}_3 \cdot \vec{x}_3 )] \, \cdot 
 \nonumber \\
 & &
\Big\{ \;
 \bar{u}^{(1)} (\vec{p}_1,s_1,t_1,c_1) \,
 \bar{u}^{(2)} (\vec{p}_2,s_2,t_2,c_2) \,
 \bar{u}^{(3)} (\vec{p}_3,s_3,t_3,c_3) \nonumber \\
 & &
 < 0 | \, T \left( \, \psi^{\,(1)} (x_1) \, \psi^{\,(2)} (x_2) \, \psi^{\,(3)}
 (x_3) \, \right) \, |P,p^+>
  \nonumber \\
 & &
 b^+ (\vec{p}_3,s_3,t_3,c_3) \; 
 b^+ (\vec{p}_2,s_2,t_2,c_2) \;
 b^+ (\vec{p}_1,s_1,t_1,c_1) \;
 |0> \; - \;
\sum\limits_{s_4,s_5} \;
\sum\limits_{t_4,t_5} \;
\sum\limits_{c_4,c_5} 
\nonumber \\
& & 
\int 
\frac{d^{\, 3}p_4}{(2\,\pi )^3 \; 2\, m_4} \; 
\frac{d^{\, 3}p_5}{(2\,\pi )^3 \; 2\, m_5} \; 
\int 
d^{\, 3}x_4 \;
d^{\, 3}x_5 \;
 \exp[ -i\, (
 \vec{p}_4 \cdot \vec{x}_4 
 + \vec{p}_5 \cdot \vec{x}_5 )] \cdot
 \nonumber \\
 & & \nonumber \\
 & & 
 \bar{u}^{(1)} (\vec{p}_1,s_1,t_1,c_1) \,
 \bar{u}^{(2)} (\vec{p}_2,s_2,t_2,c_2) \,
 \bar{u}^{(3)} (\vec{p}_3,s_3,t_3,c_3) \,
 \bar{u}^{(4)} (\vec{p}_4,s_4,t_4,c_4) \nonumber \\
 & &
 \quad < 0 | \, T \left( \, \psi^{\, (1)} (x_1) \, \psi^{\, (2)} (x_2) 
 \, \psi^{\, (3)} (x_3) \, \psi^{\, (4)} (x_4)
 \,\bar{\psi}^{\, (5)} (x_5) \, \right) \, |P,p^+>
\cdot \nonumber \\
 & & 
 \bar{v}^{(5)} (\vec{p}_5,s_5,t_5,c_5) 
 \;
 d^+ (\vec{p}_5,s_5,t_5,c_5) \; 
 b^+ (\vec{p}_4,s_4,t_4,c_4) \; 
 b^+ (\vec{p}_3,s_3,t_3,c_3) \nonumber \\
 & &
 b^+ (\vec{p}_2,s_2,t_2,c_2) \;
 b^+ (\vec{p}_1,s_1,t_1,c_1) \;
 |0> +
 \ldots  \quad \Big\} {\Big|}_{x^0_1\;=\;x^0_2 \;=\;\ldots \;=\;0}  
\end{eqnarray}
\begin{eqnarray}<K^+(\vec{P})|K^+(\vec{P}^{\;\prime})> \; = &  
<P,K^+|P^\prime,K^+> & = \; (2\pi )^3 \; 2 \, 
\omega_K (|\vec{P}\;|) \; 
\delta^3 (\vec{P}-\vec{P}^{\;\prime}) \nonumber \\
<p^+(\vec{P})|\,p^+(\vec{P}^{\;\prime})> \; = &  
<P,p^+|P^\prime,p^+> & = \; (2\pi )^3 \; \; 2 \, 
\omega_p (|\vec{P}\;|) \; \delta^3 (\vec{P}-\vec{P}^{\;\prime}) \nonumber 
\end{eqnarray}
A similar expression holds for the outgoing $\Lambda$--Baryon state 
$|\Lambda (P) >$. Both, the proton and the $\Lambda$ are considered to be
quark--diquark objects. Obviously the five quark BS--amplitude above
is connected to the ``strangeness content'' of the proton.
As for the baryons $p^+$ and $\Lambda$ the BS--amplitude of the $K^+$ is calculated via a separable
BS--kernel. The further evaluation of equ.\ (\ref{xlb1}) and (\ref{xlb2}) is performed via Wick's
theorem. Final integrations are performed in lightcone variables. A similar approach to $pp\rightarrow pp\,\phi$ (quark--gluon picture)
and $pp\rightarrow d\pi^+$ (nucleon--meson picture) at threshold is on the way. 

\end{document}